\overfullrule=0pt
\input harvmac
\def\a{{\alpha}}
\def\ad{{\dot a}}

\def\bd{{\dot b}}
\def\l{{\lambda}}
\def\lb{{\overline\lambda}}
\def\b{{\beta}}

\def\g{{\gamma}}

\def\d{{\delta}}
\def\e{{\epsilon}}
\def\s{{\sigma}}
\def\k{{\kappa}}

\def\half{{1\over 2}}
\def\p{{\partial}}

\def\t{{\theta}}
\def\tb{{\overline\theta}}
\def\bar{\overline}

\Title{\vbox{\hbox{IFT-P.031/2005 }}}
{\vbox{
\centerline{\bf Pure Spinor Formalism as an N=2 Topological
String}}}
\bigskip\centerline{Nathan Berkovits\foot{e-mail: nberkovi@ift.unesp.br}}
\bigskip
\centerline{\it Instituto de F\'\i sica Te\'orica, Universidade Estadual
Paulista}
\centerline{\it Rua Pamplona 145, 01405-900, S\~ao Paulo, SP, Brasil}

\vskip .3in
Following suggestions of Nekrasov and Siegel, a non-minimal set of fields are
added to the pure spinor formalism for the superstring. Twisted
$\hat c=3$ N=2 generators are then constructed where the pure
spinor BRST operator is the fermionic spin-one generator, and
the formalism is interpreted as a critical
topological string. Three applications of this topological string theory
include the super-Poincar\'e covariant computation of multiloop
superstring amplitudes without picture-changing operators, the construction
of a cubic open superstring field theory without contact-term problems,
and a new four-dimensional version of the pure spinor formalism which
computes F-terms in the spacetime action.

\vskip .3in

\Date {September 2005}

\newsec{Introduction}

Five years ago, a new formalism for the superstring was proposed which
is manifestly super-Poincar\'e covariant and which can be easily quantized
\ref\pureone{N. Berkovits, {\it Super-Poincar\'e covariant quantization of the
superstring}, JHEP 04 (2000) 018, hep-th/0001035.}\ref\puretwo
{N. Berkovits, {\it ICTP lectures on covariant quantization of the
superstring}, hep-th/0209059.}.
The main new feature of the formalism is a BRST operator
$Q=\int dz \l^\a d_\a$ constructed from the fermionic Green-Schwarz constraint
$d_\a$ and a bosonic ghost $\l^\a$ satisfying the pure spinor constraint
$\l^\a\g^m_{\a\b}\l^\b=0$. 
This super-Poincar\'e covariant formalism has had various 
applications such as quantization of the superstring in an $AdS_5\times S^5$ 
Ramond-Ramond background \ref\ads{N. Berkovits and O. Chand\'{\i}a, 
{\it Superstring vertex operators in an $AdS_5\times S^5$ background},
Nucl. Phys. B596 (2001) 185, hep-th/0009168.}
and computation of multiloop scattering amplitudes \ref\loop{N. Berkovits,
{\it Multiloop amplitudes and vanishing theorems using the
pure spinor formalism for the superstring}, 
JHEP 0409 (2004) 047, hep-th/0406055.}.

Because of the simple but unconventional form of the BRST operator, it is
not obvious how it can be obtained by gauge-fixing a 
reparameterization-invariant worldsheet action. Although the matter sector 
of the formalism involves the standard Green-Schwarz-Siegel 
worldsheet variables,
the ghost sector is lacking the usual $(b,c)$ ghosts and involves a 
constrained bosonic ghost $\l^\a$ with ghost-number anomaly $-8$
whose complex conjugate is absent from
the formalism. In this paper, these mysterious features of the pure spinor
formalism will be explained.

Following suggestions of Nekrasov and Siegel, 
a non-minimal set of variables which include the 
complex conjugate to $\l^\a$ and a fermionic constrained spinor
are added to the pure spinor formalism. These non-minimal variables do
not affect the BRST cohomology but change the ghost-number anomaly
from $-8$ to $+3$. The new variables
are closely related to the 
variables used for $(\beta,\gamma)$ systems
in the N=(0,2) models discussed in 
\ref\Wit{E. Witten, {\it Two-dimensional models with (0,2) supersymmetry:
perturbative aspects}, hep-th/0504078.}.
A twisted set of $\hat c=3$ N=2
superconformal generators are then constructed out of the non-minimal
variables such that
the pure spinor BRST operator is the fermionic spin-one generator.
This $\hat c=3$ N=2 superconformal field theory is then interpreted
as a critical topological string
\ref\cs{E. Witten,
{\it Chern-Simons gauge theory as a string theory}, Prog. Math. 133 (1995)
637, hep-th/9207094.}
\ref\kodaira{M. Bershadsky, S. Cecotti, H. Ooguri and
C. Vafa, {\it Kodaira-Spencer theory of gravity and
exact results for quantum string amplitudes}, Commun. Math. Phys. 165 (1994)
311, hep-th/9309140.}
in which the fermionic spin-two
generator plays the role of the $b$ ghost.

In this topological string interpretation
of the pure spinor formalism,
the simple form of the BRST operator and the absence of fundamental
$(b,c)$ ghosts
are naturally explained.
Furthermore, it will be possible to
apply standard topological string methods to compute super-Poincar\'e
covariant multiloop 
superstring amplitudes, construct a cubic superstring field theory action, and 
compactify the pure spinor formalism to four dimensions.

Using the old ``minimal'' version of the pure spinor formalism, a multiloop
amplitude prescription involving picture-changing operators was proposed
in \loop. Because the picture-changing operators required choices of constant
spacetime spinors and tensors, this prescription was only Lorentz-covariant
up to BRST-trivial surface
terms. Using the new ``non-minimal'' version of the pure spinor formalism,
multiloop superstring amplitudes can now be computed using
topological string methods in which the picture-changing operators are
replaced by a regularization factor for the zero modes.
This ``non-minimal'' prescription is manifestly Lorentz-covariant
and is expected to reproduce the ``minimal'' prescription in
a gauge in which the contribution from the non-minimal fields
decouple.

Since the superstring amplitude prescription no longer requires
picture-changing operators, the analogous open superstring field theory
action does not require singular insertions at the midpoint. 
Using standard topological
methods, one can therefore construct a cubic open
superstring field theory action resembling the Chern-Simons action \cs\ 
which does not suffer from contact-term or gauge invariance problems.
Construction
of a similar action was attempted four years ago by Schwarz and Witten
\ref\schw{J. Schwarz and E. Witten, private communication.},
but was abandoned because of difficulties caused by the ``minimal''
pure spinor measure factor. It would be interesting to generalize
this construction to a closed superstring field theory action
which might resemble the Kodaira-Spencer action \kodaira.

Critical topological strings describe Calabi-Yau compactifications
to four dimensions \kodaira\ref\antoniadis{I. Antoniadis, E. Gava, 
K.S. Narain and T.R. Taylor, {\it Topological amplitudes in string
theory}, Nucl. Phys. B413 (1994) 162, hep-th/9307158.}, 
so it is natural to consider a four-dimensional version
of the pure spinor formalism in which $\l^a$ is a $d=4$ pure spinor,
i.e. a two-component
chiral spinor. After including the 
$(x^m,\t^a,\tb^{\dot a},p_a,\bar p_\ad)$ 
variables of N=1 $d=4$ superspace, as well as
the appropriate non-minimal variables, one finds that the $d=4$ version
of the pure spinor formalism has $\hat c=0$. So after adding an N=2
$\hat c=3$ sector for the Calabi-Yau variables, one obtains a critical
topological string with manifest $d=4$ super-Poincar\'e invariance. 
But unlike the $d=4$ hybrid formalism \ref\hybrid
{N. Berkovits, {\it Covariant quantization of the Green-Schwarz
superstring in a Calabi-Yau background}, Nucl. Phys. B431 (1994) 258,
hep-th/9404162\semi N. Berkovits, {\it A new description of the superstring},
hep-th/9604123.}
which is related
to the RNS formalism by a field redefinition and describes the complete
superstring, this new formalism only describes
the chiral sector of $d=4$ superstring theory.
Note that unlike in $d=10$, $Q=\l^a d_a$ has trivial cohomology in $d=4$,
so the four-dimensional pure spinor formalism cannot be used to compute
generic superstring amplitudes. Nevertheless, the formalism
can be used to compute F-terms
in the spacetime action, and can be understood as a $d=4$
super-Poincar\'e covariant version of the $\hat c=5$ topological string
introduced in \ref\oog{
N. Berkovits, H. Ooguri and C. Vafa, {\it 
On the worldsheet derivation of large N dualities for the superstring},
Commun. Math. Phys. 252 (2004) 259, hep-th/0310118.}. 
Hopefully, this new four-dimensional formalism will
be useful for studying the effect of Ramond-Ramond fields on 
the spacetime superpotential.

In earlier papers, there have been various proposals for a more ``geometric''
version of the pure spinor formalism, some of which share certain
properties with
the non-minimal pure spinor formalism presented here.
For example, one proposal suggests relaxing the pure spinor
constraint and adding ghosts-for-ghosts to the formalism which allows
N=2 worldsheet supersymmetry
\ref\grassitwo{P.A. Grassi, G. Policastro and P. van
Nieuwenhuizen,{\it The quantum superstring as a WZNW model with N=2
superconformal symmetry}, Nucl. Phys. B676 (2004) 43.}. 
However, the N=2 worldsheet supersymmetry transformations in
this proposal are quite different from
the N=2 transformations in the non-minimal
pure spinor formalism, and the
ghosts-for-ghosts do not play the role of
non-minimal fields since they affect the 
BRST cohomology.

Another proposal has been to obtain the pure spinor formalism
from an extended Green-Schwarz formalism which involves an additional fermionic
spinor variable 
\ref\Kaz{Y. Aisaka and Y. Kazama, {\it Origin of pure spinor superstring},
JHEP 0505 (2005) 046.}
\ref\march{N. Berkovits and D.Z. Marchioro,
{\it Relating the Green-Schwarz and pure spinor formalisms
for the superstring}, JHEP 0501 (2005) 018.}
\ref\Gaona{A. Gaona and J.A. Garcia, {\it BFT embedding of the Green-Schwarz
superstring and the pure spinor formalism.}, hep-th/0507076.}. 
Unfortunately, the pure spinor BRST operator is obtained in
this proposal by passing through a complicated procedure which has up to now
only been defined in semi-light-cone gauge. Since the structure of the
worldsheet ghosts and supermoduli in semi-light-cone gauge is not well
understood, this proposal has not yet shed much light on 
the pure spinor formalism. Nevertheless, it is interesting that the
non-minimal pure spinor formalism also involves
an additional fermionic spinor variable. 

A third proposal has been to relate the pure spinor formalism to 
an N=2 super-embedding
of the Green-Schwarz superstring
\ref\superemb{M. Matone, L. Mazzucato, I. Oda, D. Sorokin and M. Tonin,
{\it The superembedding origin of the Berkovits pure spinor 
covariant quantization
of superstrings}, Nucl. Phys. B639 (2002) 182, hep-th/0206104.}, 
also known as the N=2 twistor-string \ref\tonintwo{M. Tonin,
{\it Worldsheet supersymmetric formulations of Green-Schwarz
superstrings}, Phys. Lett. B266 (1991) 312.}, and
to the $d=4$ hybrid formalism
\ref\chand{O. Chand\'{\i}a, private communication.}. 
Although the N=2 twistor-string has only
been covariantly studied at the classical level, it can be
quantized in a U(4)-covariant manner \ref\ufour{N. Berkovits,
{\it The heterotic Green-Schwarz superstring on an N=(2,0)
superworldsheet}, Nucl. Phys. B379 (1992) 96,
hep-th/9201004.} and related to the hybrid
formalism for the superstring which has $\hat c=2$ \ref\twistedrns
{N. Berkovits, {\it The ten-dimensional Green-Schwarz superstring is
a twisted Neveu-Schwarz-Ramond string}, Nucl. Phys. B420 (1994) 332,
hep-th/9308129.}\ref\topovafa{N. Berkovits and C. Vafa, {\it N=4
topological strings}, Nucl. Phys. B433 (1995) 123, hep-th/9407190.}.
Despite the fact that the N=2 twistor-string and hybrid formalism 
have different central charge from the non-minimal pure spinor formalism,
the classical N=2 worldsheet supersymmetry transformations
are very similar in the formalisms. 
It would be very interesting to understand
the relation between the $\hat c=3$ non-minimal pure spinor formalism
which describes a critical topological N=2 string and the $\hat c=2$
hybrid formalism which describes a critical non-topological N=2 string.

There have also been papers which expand on the analogy
with Chern-Simons in \puretwo \ref\superp{N. Berkovits,
{\it Covariant quantization of the superparticle using pure
spinors}, JHEP 0109 (2001) 016, hep-th/0105050.}
to find various topological properties of
the pure spinor formalism
\ref\schern{P. Grassi and G. Policastro, {\it Super-Chern-Simons
theory as superstring theory}, hep-th/0412271.}
\ref\schwarz{M. Movshev and A. Schwarz, {\it On maximally supersymmetric
Yang-Mills theories}, Nucl. Phys. B681 (2004) 324, hep-th/0311132.}
\ref\movsh{M. Movshev, {\it Yang-Mills theories in dimension
3,4,6,10 and Bar-duality}, ~~~~~ hep-th/0503165.}
\ref\tontop{I. Oda and M. Tonin, {\it Worldline approach of 
topological BF theory}, hep-th/0506054.}.
These topological properties include the construction
of the Batalin-Vilkovisky 
action, the role of the pure spinor measure factor,
and the geometrical interpretation of picture-changing
operators in amplitude computations.

Finally, there have been versions of the pure spinor formalism which
involve additional fields such as the $Y$-formalism \ref\yform{I. Oda
and M. Tonin, {\it Y-formalism in pure spinor quantization of 
superstrings}, hep-th/0505277.} and
a pure spinor version \ref\u5{R. Roiban,
W. Siegel and D. Vaman, unpublished.} of the ``Big Picture'' formalism
\ref\bigp{N. Berkovits, M.T. Hatsuda and W. Siegel, {\it The big
picture}, Nucl. Phys. B371 (1992) 434, hep-th/9108021.}. 
Although the additional
fields in these two approaches share some properties with 
the non-minimal fields used here, it is the $N=(0,2)$ model proposed
by Nekrasov \ref\nprivate{N. Nekrasov, private communication.}
for the $(\l^\a,w_\a)$ ghosts
of the pure spinor formalism
which most closely resembles the non-minimal formalism of
this paper.

In section 2 of this paper, the ``minimal'' pure spinor formalism 
will be reviewed. In section 3, a set of
``non-minimal'' variables will be added to the formalism and twisted
$\hat c=3$ N=2 generators will be constructed. 
In section 4, this critical topological string wil be used to compute
superstring scattering amplitudes up to two loops. 
In section 5, a consistent cubic open superstring field theory 
action will be constructed.
In section 6, a new four-dimensional version of the pure spinor
formalism will be defined which computes F-terms in the spacetime action.
And in the appendix, the constrained variables of the non-minimal
pure spinor formalism will be solved in terms of U(5)-covariant free
fields.

\newsec{Review of Minimal Pure Spinor Formalism}

\subsec{Worldsheet variables}

As in Siegel's approach to the Green-Schwarz superstring
\ref\siegel{W. Siegel, {\it Classical superstring mechanics}, Nucl. Phys.
B263 (1986) 93.}, the pure spinor
formalism for the superstring is constructed using the $(x^m,\t^\a)$
variables of $d=10$ superspace where $m=0$ to 9 and $\a=1$ to 16, 
together with the fermionic conjugate momenta $p_\a$. Furthermore, one
introduces a bosonic spinor ghost $\l^\a$ 
which satisfies the pure spinor constraint 
\eqn\puredef{\l^\a \g^m_{\a\b}\l^\b=0}
where $\g^m_{\a\b}$ are the symmetric $16\times 16$ $d=10$ Pauli matrices.

Because of the pure spinor constraint on $\l^\a$, its
conjugate momentum $w_\a$ is defined up
to the gauge transformation 
\eqn\gaugeone{\d w_\a = \Lambda^m (\g_m\l)_\a,}
which implies that $w_\a$ only appears through its Lorentz current
$N_{mn}$, ghost current $J_\l$, and
stress tensor $T_\l$. These gauge-invariant currents are defined by
\eqn\currone{N_{mn}=\half w\g_{mn}\l,\quad J_\l = w_\a\l^\a,\quad
T_\l = w_\a \p\l^\a.}

The worldsheet action for the left-moving matter and ghost variables is 
\eqn\actionpure{S=\int d^2 z (\half \p x^m \bar\p x_m + p_\a \bar\p\t^\a
- w_\a \bar\p\l^\a),}
and the right-moving variables will be ignored throughout this paper.
For the Type II superstring, the right-moving variables are similar to
the left-moving variables, while for the heterotic superstring, the
right-moving variables are the same as in the RNS heterotic formalism.

The OPE's for the matter variables are easily computed to be
\eqn\matterope{x^m (y) x^n(z)\to -\eta^{mn}\log|y-z|^2, \quad
p_\a(y)\t^\b(z) \to \d_\a^\b (y-z)^{-1},}
however, the pure spinor constraint on $\l^\a$ prevents a direct computation
of the OPE's for the ghost variables. Nevertheless, one can compute OPE's
involving $\l^\a$ and the currents of \currone\ either by solving the 
pure spinor constraint in terms of U(5)-covariant free fields \pureone, 
by using
the SO(10)-covariant fixed-point techniques of \ref\nikita
{N. Berkovits and N. Nekrasov, {\it Character of pure spinors},
hep-th/0503075.}, or by using the
$Y$-formalism of \yform. The resulting OPE's are
\eqn\ope{ N_{mn}(y) \l^\a(z) \to \half (y-z)^{-1} (\gamma_{mn}\l)^\a, \quad
J(y) \l^\a(z) \to (y-z)^{-1} \l^\a,}
$$N^{kl}(y) N^{mn}(z) \to 
- 3 (y-z)^{-2}
(\eta^{n[k} \eta^{l]m}) +
(y-z)^{-1}(\eta^{m[l} N^{k]n} -
\eta^{n[l} N^{k]m} ) 
,$$
$$ J_\l(y) J_\l(z) \to -4 (y-z)^{-2}, \quad J_\l(y) N^{mn}(z) 
\to {\rm regular}, $$
$$N_{mn}(y) T_\l(z) \to (y-z)^{-2} N_{mn}(z) ,\quad 
J_\l(y) T_\l(z) \to  -8(y-z)^{-3} + (y-z)^{-2} J_\l(z),$$
$$T_\l(y) T_\l(z) \to 11(y-z)^{-4}  + 2(y-z)^{-2} T_\l(z) + 
(y-z)^{-1}\p T_\l(z).$$

{}From the above OPE's, one sees that the central charge contribution
to the conformal anomaly is $22$, the level for the Lorentz currents is
$-3$, and the ghost-number anomaly is $-8$. So the central charge 
contribution from the ghost variables cancels the contribution of 
$+10-32=-22$ from the $(x^m,\t^\a,p_\a)$ matter variables. 
Furthermore, the total Lorentz current is $M_{mn}= -\half(p\g_{mn}\t) + 
N_{mn}$,
and since $-\half(p\g_{mn}\t)$ has level $+4$, $M_{mn}$ has
the same level of $+4-3=1$ as the RNS Lorentz current $M_{mn}=\psi_m\psi_n$.
Finally, it will be explained in the following section that after
adding a set of non-minimal variables, the
ghost-number anomaly of $-8$ is shifted to the usual ghost-number anomaly
of $+3$.

\subsec{Physical states}

Physical open string states in the pure spinor formalism are defined
as ghost-number one states in the cohomology of the nilpotent BRST
operator
\eqn\purebrst{Q=\int dz ~\l^\a d_\a}
where 
\eqn\ddef{d_\a= p_\a -\half\g_{\a\b}^m \t^\b \p x_m -{1\over 8}
\g_{\a\b}^m \g_{m\g\d}\t^\b\t^\g\p\t^\d}
is the supersymmetric Green-Schwarz constraint.
As shown by Siegel \siegel, $d_\a$ satisfies the OPE's 
\eqn\oped{d_\a(y) d_\b(z) \to -(y-z)^{-1} \g_{\a\b}^m \Pi_m,\quad 
d_\a(y) \Pi^m(z) \to  (y-z)^{-1} \g_{\a\b}^m \p\t^\b(z),}
$$d_\a(y)~ f(x(z),\t(z))\to (y-z)^{-1} D_\a f(x(z),\t(z)),$$
where 
\eqn\susyd{D_\a=
{\p\over{\p\t^\a}} +\half \t^\b \g^m_{\a\b} \p_m }
is the $d=10$ supersymmetric
derivative, 
$\Pi^m = \p x^m +\half\t \g^m \p\t$
is the supersymmetric momentum and 
\eqn\defqq{
q_\a= \int dz (p_\a +\half\g_{\a\b}^m \t^\b \p x_m +{1\over {24}}
\g_{\a\b}^m \g_{m~\g\d}\t^\b\t^\g\p\t^\d)}
is the supersymmetric generator satisfying
\eqn\sus{ \{q_\a,q_\b\}=
\g_{\a\b}^m\int dz \p x_m, \quad [q_\a, \Pi^m(z)]=0,\quad
\{q_\a, d_\b(z)\}=0.}

For massless states described by $V=\l^\a A_\a(x,\t)$, $QV=0$ and
$\d V= Q\Omega$ implies that $A_\a$ is the super-Yang-Mills spinor
gauge field satisfying the linearized
equation of motion $ (\g_{mnpqr})^{\a\b} D_\a
A_\b=0$ and the linearized
gauge invariance $\d A_\a = D_\a \Omega$. For massive
states, the superspace description is more complicated 
\ref\massivec{N. Berkovits and O. Chandia, {\it Massive superstring
vertex operator in D=10 superspace}, JHEP 0208 (2002) 040,
hep-th/0204121.}, however, it
has been proven by DDF methods that the cohomology of $Q$ at ghost-number
one correctly
describes the open superstring spectrum \ref\cohomo{N. Berkovits,
{\it Cohomology in the pure spinor formalism for the superstring},
JHEP 09 (2000) 046, hep-th/0006003.}.

\subsec{Scattering amplitudes}

To compute scattering amplitudes using the ``minimal'' pure spinor
formalism, it is necessary to introduce picture-changing operators which
can absorb the zero modes of the bosonic ghosts $\l^\a$ and $w_\a$.
For example, $N$-point tree amplitudes are computed by the correlation 
function
\eqn\corrtreemin{{\cal A} = \langle V_1(z_1) V_2(z_2) V_3(z_3)
\int dz_4 U_4(z_4) ... \int dz_N U_N(z_N)~\prod_{I=1}^{11}
Y_{C_I}(y_I) \rangle}
where $Y_{C_I} = C_{I\a}\t^\a \d(C_{I\b}\l^\b)$ are
picture-lowering operators which absorb the eleven $\l^\a$ zero modes,
$C_{I\a}$ are constant spinors,
and $U_r$ are dimension-one vertex operators which are related to
the unintegrated vertex operators $V_r$ by the relation $Q U_r=\p V_r$.
This tree amplitude prescription has been shown to coincide with the
RNS prescription for massless states with an arbitrary number of
bosons and up to four fermions \ref\vali{N. Berkovits and B.C. Vallilo,
{\it Consistency of super-Poincar\'e covariant superstring tree amplitudes},
JHEP 07 (2000) 015, hep-th/0004171.}.

$N$-point $g$-loop amplitudes can also be computed in the minimal
pure spinor formalism by evaluating the correlation function
\eqn\corrthreemin{{\cal A} = \int d^{3g-3}\tau \langle 
\prod_{j=1}^{3g-3}(\int dw_j \mu_j(w_j) \tilde b_{B_j}(w_j)) 
\prod_{P=3g-2}^{10g}Z_{B_P}(w_P) \prod_{R=1}^g Z_J(v_R) }
$$\prod_{I=1}^{11} Y_{C_I}(y_I) \prod_{r=1}^N 
\int dz_r U(z_r)  \rangle $$
where $\tau_j$ are complex Teichmuller parameters and $\mu_j$ are
the associated Beltrami differentials, 
$Z_B= B_{mn} (\l\g^{mn}d) \d(B_{mn} N^{mn})$ and $Z_J= (\l^\a d_\a) \d(J_\l)$
are picture-raising operators which absorb the $11g$ zero modes of $w_\a$,
$B_{mn}$ are constant tensors,
and $\tilde b_B$ is a picture-raised 
$b$ ghost which is defined to satisfy $\{Q, \tilde b_B\} = T Z_B$.
Although the explicit form of $\tilde b_B$ is quite complicated, this
amplitude prescription has been used to prove various vanishing theorems
and to compute four-point one-loop and two-loop massless amplitudes \loop
\ref\vanh{L. Anguelova, P.A. Grassi and P. Vanhove,
{\it Covariant one-loop amplitudes in D=11}, Nucl. Phys. B702 (2004) 269,
hep-th/0408171.}\ref\twoloop{N. Berkovits, {\it Super-Poincar\'e covariant
two-loop superstring amplitudes}, ~~~~~ hep-th/0503197.}.

Although the choices of constant spinors $C_\a$ and tensors $B_{mn}$ 
in the picture-changing operators $Y_C$ and $Z_B$ break manifest
Lorentz covariance, one can show that the dependence on
$C_\a$ and $B_{mn}$ is BRST-trivial. So after integrating over the
Teichmuller parameters, the scattering amplitude is independent of
the choices for $C_\a$ and $B_{mn}$. Nevertheless, it would be more
convenient if Lorentz covariance could be manifestly preserved at
all stages in the
amplitude computation. As will now be shown, this is possible using
a ``non-minimal'' version of the pure spinor formalism in which 
picture-changing
operators are replaced by a regularization factor for the zero modes.

\newsec{Non-Minimal Pure Spinor Formalism}

\subsec{Worldsheet variables}

Although the BRST operator in the pure spinor formalism has a simple
structure, the lack of a geometrical
interpretation of the formalism makes
it difficult to understand the rules for computing scattering amplitudes.
As will be explained here, after introducing a set of non-minimal variables,
the pure spinor formalism can be interpreted as a critical topological string
with the standard topological rules for computing scattering amplitudes.

The new non-minimal variables will consist of a bosonic pure spinor
$\bar\lambda_\a$ and 
a constrained fermionic spinor $r_\a$ satisfying the constraints
\eqn\newcons{\lb_\a\g_m^{\a\b}\lb_\b=0 \quad {\rm and}\quad
\lb_\a \g_m^{\a\b} r_\b=0.}
In d=10 Euclidean space where complex conjugation flips the chirality
of spacetime spinors, $\lb_\a$ can be interpreted as the complex conjugate
to $\l^\a$.
The worldsheet action for the non-minimal pure spinor formalism is
\eqn\nonmin{\int d^2 z (\half \p x^m \bar\p x_m +p_\a \bar\p\t^\a
- w_\a\bar\p \l^\a - \bar w^\a \bar\p \bar\l_\a + s^\a \bar\p r_\a )}
where $\bar w^\a$ and $s^\a$ are the conjugate momenta for $\bar\l_\a$
and $r_\a$ with $+1$ conformal weight. As explained in the appendix,
the constraints of \newcons\ can be solved in a U(5)-covariant
manner and $\lb_\a$ and $r_\a$
can be expressed in terms of eleven independent bosonic and fermionic free
fields. Note that all non-minimal variables
are left-moving on the worldsheet (like $\l^\a$ and $\t^\a$), and that
$\lb_\a$ and $r_\a$ are spacetime spinors of opposite chirality from
$\l^\a$ and $\t^\a$. It is interesting that similar variables to
$\lb_\a$ and $r_\a$ have recently been used in N=(0,2) models for
chiral $(\b,\g)$ systems \Wit. 
However, unlike in these N=(0,2) models where
the additional variables move in the opposite direction on the worldsheet
from the $(\b,\g)$ variables, the non-minimal variables in the pure
spinor formalism move in the same direction on the worldsheet
as the $(\l^\a, w_\a)$ variables.

Just as $w_\a$ can only appear in the gauge-invariant combinations
\eqn\mingauge{
N_{mn}= \half (w\g_{mn}\l),\quad J_\l=w_\a\l^\a,\quad T_\l = w_\a\p\l^\a,}
the variables $\bar w^\a$ and $s^\a$ can only appear in the combinations
\eqn\gaugeinvbar{\bar N_{mn}= \half (\bar w\g_{mn}\lb - s\g_{mn} r),\quad
\bar J_{\bar \l} =\bar w^\a \bar\l_\a -s^\a r_\a,\quad 
T_\lb = \bar w^\a\p\bar\l_\a - s^\a \p r_\a,}
$$S_{mn} = \half s \g_{mn} \lb,\quad  S=  s^\a \lb_\a,$$
which are invariant under the gauge transformations
\eqn\gaugebar{\d \bar w^\a = \bar\Lambda^m (\g_m\lb)^\a - \phi^m (\g_m r)^\a,
\quad \d s^\a = \phi^m (\g_m \lb)^\a}
for arbitrary $\bar\Lambda^m$ and $\phi^m$.
Note that $J_r =r_\a s^\a$ and $\Phi= \bar w^\a r_\a$ are
also gauge-invariant, but they can be written in terms of the
other currents as 
\eqn\other{J_r = {{(\l r) S -{2\over 3} (\l\g^{mn}r)S_{mn}}\over {(\l\lb)}},
\quad \Phi = 
{{(\l r)(\bar J_\lb + J_r) -{2\over 3}
(\l\g^{mn} r) \bar N_{mn}}\over {(\l\lb)}}.}
These gauge-invariant currents will be shown in the appendix
to satisfy the OPE's
\eqn\opebar{ \bar N_{mn}(y) \lb_\a(z) \to 
\half (y-z)^{-1} (\gamma_{mn}\lb)_\a, 
\quad \bar N_{mn}(y) r_\a(z) \to 
\half (y-z)^{-1} (\gamma_{mn}r)_\a, }
$$
\bar J_\lb (y) \lb_\a(z) \to (y-z)^{-1} \lb_\a,\quad
J_\lb (y) r_\a(z) \to (y-z)^{-1} r_\a,$$
$$\bar N^{kl}(y) \bar N^{mn}(z) \to 
(y-z)^{-1}(\eta^{m[l} 
\bar N^{k]n} -
\eta^{n[l} \bar N^{k]m} ) ,$$
$$\bar J_\lb(y) \bar N^{mn}(z) 
\to {\rm regular},
\quad  J_r(y) \bar N^{mn}(z) 
\to {\rm regular}, 
\quad  \Phi(y) \bar N^{mn}(z) 
\to {\rm regular}, $$
$$\Phi(y)\lb_\a(z)\to (y-z)^{-1}r_\a,\quad
\Phi(y) S_{mn}(z)\to (y-z)^{-1}\bar N_{mn},\quad
\Phi(y) S(z)\to (y-z)^{-1}\bar J_\lb,$$
$$ \bar J_\lb(y) \bar J_\lb(z) \to {\rm regular}, \quad
J_r(y) J_r(z) \to 11 (y-z)^{-2}, 
\quad J_\lb (y) J_r(z) \to 8 (y-z)^{-2},$$
$$\bar N_{mn}(y) T_\lb(z) \to (y-z)^{-2} \bar N_{mn}(z) ,$$
$$ 
\bar J_\lb(y) T_\lb(z) \to   (y-z)^{-2}\bar J_\lb(z), $$
$$ J_r(y) T_\lb(z) \to  11(y-z)^{-3} + (y-z)^{-2} J_r(z),$$
$$T_\lb(y) T_\lb(z) \to  2(y-z)^{-2} T_\lb(z) + 
(y-z)^{-1}\p T_\lb(z).$$

{}From the above OPE's, one sees that the non-minimal
variables do not contribute 
to the conformal anomaly or to the level of the Lorentz currents.
Furthermore,
if the ghost current is defined as $w_\a\l^\a-\bar w^\a\lb_\a =
J_\l - \bar J_\lb + J_r$, 
the non-minimal variables shift
the ghost-number anomaly to $-8 + 11= +3$, which is the same
ghost-number anomaly as in bosonic string theory.

\subsec{$\hat c=3$ N=2 generators}

In order that the non-minimal variables do not affect the cohomology,
the ``minimal'' pure spinor BRST operator $Q=\int dz \l^\a d_\a$ will
be modified to the ``non-minimal'' BRST operator \nprivate
\eqn\nonminQ{Q_{nonmin} = \int dz (\l^\a d_\a + \bar w^\a r_\a).}
The new term $\int dz \bar w^\a r_\a$ is invariant under the gauge
transformation of \gaugebar\
and implies through the usual quartet argument that
the cohomology is independent of $(\bar \l_\a, \bar w^\a)$ and
$(r_\a, s^\a)$.

In the ``minimal'' pure spinor formalism, one could have 
defined a non-covariant
$b$ ghost satisfying $\{Q, b\}= T$  as 
\ref\rns{N. Berkovits, {\it Relating the RNS and
pure spinor formalisms for the superstring}, JHEP 08 (2001) 026,
hep-th/0104247.}
\eqn\minb{b = {{C_\a G^\a}\over{C_\a \l^\a}}}
where $C_\a$ is any constant spinor and 
\eqn\defG{G^\a 
=\half\Pi^m (\g_m d)^\a -{1\over 4} N_{mn}(\g^{mn}\p\t)^\a
-{1\over 4} J_\l \p\t^\a -{1\over 4} \p^2\t^\a}
satisfies $\{Q, G^\a\}=\l^\a T$.
However, such a $b$ ghost contains poles when
$C_\a \l^\a=0$, which causes problems in the presence of picture-changing
operators containing factors of $\d(\l)$.

In the non-minimal pure spinor
formalism, there will be no picture-changing operators
and one can define a Lorentz-invariant $b_{nonmin}$ ghost satisfying
$\{Q_{nonmin}, b_{nonmin} \} = T_{nonmin}$ as
\eqn\nonminb{b_{nonmin} =  s^\a\p\lb_\a  + {{\lb_\a G^\a}\over{(\lb\l)}} +
{{\lb_\a r_\b H^{[\a\b]}}\over{(\lb\l)^2}} 
 -
{{\lb_\a r_\b r_\g K^{[\a\b\g]}}\over{(\lb\l)^3}} -
{{\lb_\a r_\b r_\g r_\d L^{[\a\b\g\d]}}\over{(\lb\l)^4}} }
$$=
s^\a\p\lb_\a  + {{\lb_\a (2
\Pi^m (\g_m d)^\a-  N_{mn}(\g^{mn}\p\t)^\a
- J_\l \p\t^\a -{1\over 4} \p^2\t^\a)}\over{4(\lb\l)}} $$
$$+ {{(\lb\g^{mnp} r)(d\g_{mnp} d +24 N_{mn}\Pi_p)}\over{192(\lb\l)^2}} 
-
{{(r\g_{mnp} r)(\lb\g^m d)N^{np}}\over{16(\lb\l)^3}} +
{{(r\g_{mnp} r)(\lb\g^{pqr} r) N^{mn} N_{qr}}\over{128(\lb\l)^4}} $$
where
\eqn\Tnonmin{T_{nonmin}= -\half \p x^m \p x_m - p_\a \p\t^\a + w_\a \p\l^\a
+ \bar w^\a \p\lb_\a - s^\a \p r_\a,}
and
$(G^\a,H^{\a\b},K^{\a\b\g},L^{\a\b\g\d})$ are operators
which were defined in \loop\yform\
for constructing the picture-raised $\tilde b_B$ ghost
and satisfy
\eqn\ghkl{\{Q, G^\a\} = \l^\a T, \quad [Q,H^{[\a\b]}]=\l^{[\a} G^{\b]},\quad
\{Q,K^{[\a\b\g]}\}=\l^{[\a} H^{\b\g]},}
$$
[Q,L^{[\a\b\g\d]}]=\l^{[\a} K^{\b\g\d]},\quad
\l^{[\a} L^{\b\g\d\k]}=0. $$
In addition to satisfying
$\{Q_{nonmin}, b_{nonmin} \} = T_{nonmin}$,
one can verify that $b_{nonmin}$ has no poles with itself.
Note that only the antisymmetrized
components of $H^{\a\b}$,
$K^{\a\b\g}$ and $L^{\a\b\g\d}$ contribute to $b_{nonmin}$, which 
makes
the computation of coefficients in $b_{nonmin}$ much simpler than
in the computation of the picture-raised $\tilde b_B$ ghost 
\loop\ref\btonin {I. Oda and M. Tonin, 
{\it On the $b$ antighost in the pure spinor quantization of superstrings},
Phys. Let. B606 (2005) 218, hep-th/0409052.}\yform.
Although $b_{nonmin}$ appears complicated in \nonminb, 
its construction in terms
of Siegel-like constraints \siegel\ suggests that it may have a natural
superspace interpretation.

To complete the construction of the $\hat c=3$ N=2 generators, one
needs to construct the U(1) current $J_{nonmin}$ by 
computing the double pole of $b_{nonmin}$ with the integrand
of $Q_{nonmin}$. The result is
\eqn\Jnonmin{J_{nonmin} = w_\a \l^\a - s^\a r_\a 
-2 {{\lb_\a \p\l^\a + r_\a \p\t^\a}\over{(\lb\l)}} +2
 {{(\lb_\a r^\a)(\lb_\b \p\t^\b)}\over{(\lb\l)^2}} .}
The unusual non-quadratic terms in $J_{nonmin}$ can be understood
to be necessary for two reasons. Firstly, the term $(\lb_\a G^\a)/(\lb\l)$
in $b_{nonmin}$ has a double pole with $\l^\a w_\a$, which needs to be
cancelled by the double pole of $b_{nonmin}$ with the non-quadratic
terms in order that $b_{nonmin}$ is a U(1) primary field.
Secondly, the triple pole of $J_{nonmin}$ with $T_{nonmin}$ of \Tnonmin\ is 
equal to $-8 + 11 = +3$.
But the N=2 Jacobi identities imply that
this ghost-number anomaly of $+3$ should be equal to the double pole of
$J_{nonmin}$ with itself, which 
gives
the value $-4+ 11 = +7$ if one does not include the contribution from
the non-quadratic terms.

So the twisted $\hat c=3$ N=2 generators are given by 
the U(1) current $J_{nonmin}$ of \Jnonmin, the
fermionic generators $\l^\a d_\a + \bar w^\a r_\a$
and $b_{nonmin}$ of \nonminb, and the stress tensor $T_{nonmin}$ of 
\Tnonmin. Although the form of $J_{nonmin}$ is complicated, it can
be simplified by shifting by a BRST-trivial quantity as 
\eqn\Jshift{J'_{nonmin} = J_{nonmin} + \{ Q_{nonmin}, -s^\a \lb_\a
+2{{\lb_\a\p\t^\a}
\over{(\lb\l)}} \} }
$$= J_{nonmin} - \bar w^\a \lb_\a + s^\a r_\a 
+2 {{\lb_\a \p\l^\a + r_\a \p\t^\a}\over{(\lb\l)}} -2
 {{(\lb_\a r^\a)(\lb_\b \p\t^\b)}\over{(\lb\l)^2}} $$
$$ = w_\a \l^\a - \bar w^\a \lb_\a  .$$
Although $J'_{nonmin}$ has double poles with $b_{nonmin}$ and
does not have level $+3$, one
can easily check that $[\int dz J'_{nonmin}, Q_{nonmin}]= Q_{nonmin}$ and
$[\int dz J'_{nonmin}, b_{nonmin}] = - b_{nonmin}$. 
Furthermore, it will be shown in the appendix that the triple pole
of $J'_{nonmin}$ with $T_{nonmin}$ is $+3$, so the ghost-number anomaly is
preserved using $J'_{nonmin}$.
These are the only necessary conditions for the ghost current 
in critical topological string theory, as can be seen by comparing
with the ghost current of the bosonic string,
$J= bc$, which 
has double poles with the BRST current and whose level of $+1$ does
not coincide with its ghost-number anomaly of $+3$.
So there is no problem with
replacing $J_{nonmin}$ by $J'_{nonmin}$ in the definition of 
the topological string associated to the non-minimal pure spinor formalism.

In the next section, superstring scattering amplitudes will be computed
using topological methods with the 
U(1) charge $\int dz J= \int dz (w_\a\l^\a  -\bar w^\a \lb_\a)$,
the BRST operator $Q = \int dz (\l^\a d_\a + \bar w^\a r_\a)$, the
stress tensor 
$T = 
-\half \p x^m \p x_m - p_\a \p\t^\a + w_\a \p\l^\a
+ \bar w^\a \p\lb_\a - s^\a \p r_\a$,
and
the $b$ ghost
\eqn\bgh{b = 
s^\a\p\lb_\a  + {{\lb_\a 
(2\Pi^m (\g_m d)^\a - N_{mn}(\g^{mn}\p\t)^\a
- J_\l \p\t^\a - \p^2\t^\a)}\over {4(\lb\l)}} }
$$+
{{(\lb\g^{mnp} r)(d\g_{mnp} d +24 N_{mn}\Pi_p)}\over{192(\lb\l)^2}} 
-
{{(r\g_{mnp} r)(\lb\g^m d)N^{np}}\over{16(\lb\l)^3}} +
{{(r\g_{mnp} r)(\lb\g^{pqr} r) N^{mn} N_{qr}}\over{128(\lb\l)^4}} .$$
Note that for the rest of this paper, the subscript 
${nonmin}$ will be dropped
from these operators.

\newsec{ Computation of Scattering Amplitudes}

\subsec{Tree amplitudes}

Since the non-minimal pure spinor formalism is a $\hat c=3$ N=2
string theory, one can use standard methods developed for critical
topological strings to compute scattering amplitudes. For example,
$N$-point tree amplitudes are computed as in bosonic string
theory by the correlation function of three unintegrated vertex operators
$V$ satisfying $QV=0$ and $N-3$ integrated vertex operators
$\int dz U(z)$ satisfying
$QU = \p V$.
As in the minimal pure spinor formalism, functional integration over
the worldsheet variables of $+1$ conformal weight is straightforward
using the poles in the OPE's of \ope\ and
\opebar. One is then left with an expression
${\cal A} = \langle f(\l,\lb,r,\t) \rangle$ where $f(\l,\lb,r,\t)$ carries
$+3$ U(1) charge and depends only on the zero modes of $\l^\a$, $\lb_\a$,
$r_\a$ and $\t^\a$. Note that integration over the $x^m$ zero modes
is performed in the standard manner and will be ignored throughout this paper.

Since $\l^\a$ and $\lb_\a$ are non-compact bosonic variables, the
integral over the zero modes
\eqn\intone{{\cal A} = \int[d\l][d\lb][dr]d^{16}\t
 f(\l,\lb,r,\t)}
needs to be regularized. A useful regularization method developed
by Marnelius \ref\marnel{R. Marnelius and M. Ogren,
{\it Symmetric inner products for physical states in BRST quantization},
Nucl. Phys. B351 (1991) 474.}
for BRST-invariant systems involves inserting the
factor ${\cal N} = \exp (\{Q,\chi\})$ into the
integral where $\chi$ is some fermionic function of the worldsheet variables. 
Since $f(\l,\lb,r,\t)$ is BRST-invariant and
${\cal N}=1 + ...$ where ... is BRST-trivial,
the integral will
be independent of the choice of $\chi$. 

In the non-minimal pure spinor formalism, it is convenient to choose
$\chi = -\lb_\a \t^\a$ so that
\eqn\defcaln{{\cal N} = \exp(\{Q,\chi\}) = \exp (-\lb_\a\l^\a -r_\a\t^\a).}
Treating $\lb_\a$ as the complex conjugate of $\l^\a$, the expression
\eqn\inttwo{{\cal A} = \int[d\l][d\lb][dr]d^{16}\t~ {\cal N}
 f(\l,\lb,r,\t)}
is well-defined
if one assumes
that $f(\l,\lb,r,\t)$ does not diverge too fast as $\l\lb\to 0$.

To determine how fast $f(\l,\lb,r,\t)$ is allowed to diverge as $\l\lb\to 0$,
note that the measure factors $[d\l]$ and $[d\lb]$ for pure spinors 
satisfy\loop\ref\cherk{N. Berkovits and S. Cherkis, {\it Higher-dimensional
twistor transforms using pure spinors}, JHEP 0412 (2004) 049, hep-th/0409243.}
\eqn\measurelambda{[d\l] \l^\b\l^\g\l^\d = 
(\e T^{-1})_{\a_1 ... \a_{11}}^{\b\g\d}
d\l^{\a_1} ... d\l^{\a_{11}} }
and 
$$[d\lb] \lb_\b\lb_\g\lb_\d = 
(\e T)^{\a_1 ... \a_{11}}_{\b\g\d}
d\lb_{\a_1} ... d\lb_{\a_{11}} $$
where 
$(\e T^{-1})_{\a_1 ... \a_{11}}^{\b\g\d}$ and
$(\e T)^{\a_1 ... \a_{11}}_{\b\g\d}$
are Lorentz-invariant tensors defined
in \pureone\loop\ which are antisymmetric in $[\a_1 ... \a_{11}]$
and are symmetric and gamma-matrix traceless in $(\b\g\d)$. 
Up to an overall normalization constant,
\eqn\overall{
(\e T)^{\a_1 ... \a_{11}}_{\b\g\d} =
\e^{\a_1 ...\a_{16}}\g^m_{\a_{12}\rho}
\g^n_{\a_{13}\sigma}
\g^p_{\a_{14}\tau}
(\g_{mnp})_{\a_{15}\a_{16}} (\d_{(\b}^\rho \d_\g^\sigma \d_{\d)}^\tau
- {1\over{40}} \g^m_{(\b\g} \d_{\d)}^{\rho} \g_m^{\sigma\tau}).}
Furthermore, the constraint $\lb\g^m r=0$ implies that the measure
factor $[dr]$ satisfies 
\eqn\measurer{[dr] = (\e T^{-1})^{\b\g\d}_{\a_1 ... \a_{11}}
\lb_\b\lb_\g \lb_\d ({\p\over\p r_{\a_1}}) ...
({\p\over\p r_{\a_{11}}}) .}

So the measure factor $[d\l][d\lb][dr]$ goes like $\l^8 \lb^{11}$
as $\l\lb\to 0$, which implies that $f(\l,\lb,r,\t)$ must diverge
slower than $\l^{-8}\lb^{-11}$ in order that \inttwo\ is well-defined.
If one wants to compute amplitudes in which $f(\l,\lb,r,\t)$ diverges
as fast as $\l^{-8}\lb^{-11}$ when $\l\lb\to 0$, 
an alternative regularization method for
the zero modes must be found. 

The restriction that $f(\l,\lb,r,\t)$ diverges slower than $\l^{-8}\lb^{-11}$
is related to the operator $\xi = (\lb\t)/(\lb\l + r\t)$ which satisfies
$Q\xi=1$. Since $QV=0$ implies that $Q(\xi V)=V$, the existence of the
operator $\xi$ naively implies that the BRST cohomology is trivial.
For example, $f=Q(\xi f)$ where 
$f= (\l\g^m\t)(\l\g^n\t)(\l\g^p\t)(\t\g_{mnp}\t)$
naively implies that $\langle {\cal N}~f \rangle=\langle {\cal N}~
Q(\xi f)\rangle=0$.
But because of the restriction that $f$ diverges slower than 
$\l^{-8}\lb^{-11}$, $\langle {\cal N}~Q(\Omega)\rangle$ is only guaranteed to
vanish if $\Omega$ diverges slower than $\l^{-8}\lb^{-10}$ as $\l\lb\to 0$.
When 
$f= (\l\g^m\t)(\l\g^n\t)(\l\g^p\t)(\t\g_{mnp}\t)$, $\xi f$ contains terms
which diverge as $\l^{-8}\lb^{-10}(\t)^{16} (r)^{10}$. So $\xi f$ is not
an allowable gauge parameter, which explains why $\langle {\cal N}~
f\rangle \neq 0$.

So the regularized prescription for computing the $N$-point tree amplitude
using topological string methods is given by the correlation function
\eqn\corrtree{{\cal A} = \langle {\cal N}(y) V_1(z_1) V_2(z_2) V_3(z_3)
\int dz_4 U_4(z_4) ... \int dz_N U_N(z_N) \rangle}
where ${\cal N}(y) = \exp(\{Q,\chi(y)\})= \exp (-\l(y)\lb(y)-r(y)\t(y))$
and $y$ is an arbitrary point on the worldsheet. 
Suppose that
all external states are chosen in the gauge where the vertex operators
$V$ and $U$ are independent of the non-minimal fields. Then 
after integrating out the variables of $+1$ conformal weight using the
poles in their OPE's, one obtains
\eqn\obtin{{\cal A} = \langle {\cal N} f(\l,\t)\rangle = 
\langle {\cal N} \l^\a \l^\b \l^\g f_{\a\b\g}(\t)\rangle,} 
which has no divergences when $\l\lb \to 0$.
Using the measure factors defined above, one finds up to 
an overall normalization constant that
\eqn\intthree{{\cal A} = \int[d\l][d\lb][dr]d^{16}\t 
\exp (-\lb_\a\l^\a -r_\a\t^\a)
\l^\b\l^\g \l^\d f_{\b\g\d}(\t)}
$$= \int d^{16}\t (\e T^{-1})^{\b\g\d}_{\a_1 ... \a_{11}}
\t^{\a_1} ... \t^{\a_{11}} f_{\b\g\d}(\t)$$
$$
= \e^{\a_1 ... \a_{16}} 
(\e T^{-1})^{\b\g\d}_{\a_1 ... \a_{11}}
({\p\over{\p\t^{\a_{12}}}})
... ({\p\over{\p\t^{\a_{16}}}}) f_{\b\g\d}(\t),$$
which agrees with the result from the minimal pure spinor formalism.

To understand the relationship between the non-minimal and minimal
computations, note that BRST-invariance implies that the amplitude
is unaffected by rescaling $\chi=-\lb_\a\t^\a$
to $\chi= -\rho\lb_\a\t^\a$ for any positive
$\rho$ in the definition of ${\cal N}$.
So one can take the limit $\rho\to\infty$ in
${\cal N}_\rho (y)= \exp(-\rho(\l(y)\lb(y)+ r(y)\t(y)))$, which
is non-vanishing only when $\l^\a(y)=\lb_\a(y)=0$. So in the limit
$\rho\to \infty$, ${\cal N}_\rho(y)$ contains the same $\d^{11}(\l)$
dependence as the product of eleven picture-lowering operators 
$\prod_{I=1}^{11} Y_{C_I}(y)$ in the minimal formalism. However,
in addition to being manifestly Lorentz-invariant, the advantage of using
${\cal N}(y)$ instead of picture-changing operators is that one
can take the opposite limit $\rho\to 0$ in which ${\cal N}_\rho(y)$
becomes a smooth invertible function.

After introducing the regularization factor ${\cal N}= \exp(-\l\lb-r\t)$, one
can also define $N$-point tree amplitudes in a worldsheet
reparameterization invariant manner 
as
\eqn\corrtwo{{\cal A} = \langle {\cal N}(y) ~V_1(z_1) ... V_N(z_N)
\int dz_4 b(z_4) ... \int dz_N b(z_N) \rangle}
where $b(z)$ is defined in \bgh. But since each unintegrated vertex
operator $V$ goes like $\l$ and each $b$ ghost goes like
$\lb/(\lb\l)^4$, $f(\l,\lb,r,\t)$ goes like $\l^3 (\lb\l)^{9-3N}$
when $\l\lb \to 0$.
Since $f(\l,\lb,r,\t)$ must diverge slower than $\l^{-8}\lb^{-11}$,
a maximum of three $b$ ghosts (or six unintegrated vertex operators)
can be allowed in computations using this regularization method. 

\subsec{Loop amplitudes}

To compute $N$-point $g$-loop amplitudes, one uses the 
topological prescription
\eqn\corrthree{{\cal A} = \int d^{3g-3}\tau \langle {\cal N}(y)
\prod_{j=1}^{3g-3}(\int dw_j \mu_j(w_j) b(w_j)) \prod_{r=1}^N 
\int dz_r U(z_r) \rangle}
where $\tau_j$ are the complex Teichmuller parameters and $\mu_j$ are
the associated Beltrami differentials, $b(z)$
is defined in \bgh, and ${\cal N}(y)$ is a regularization
factor for the genus $g$ zero modes which will be defined below.
To define this regularization factor, first
separate off the zero modes of the gauge-invariant worldsheet
fields of $+1$ conformal weight
as
\eqn\separate{N_{mn}(z) = \widehat N_{mn}(z) + \sum_{I=1}^g N_{mn}^I 
\omega_I(z),\quad
\bar N_{mn}(z) = \widehat {\bar N}_{mn}(z) + \sum_{I=1}^g \bar N_{mn}^I 
\omega_I(z),}
$$
J_\l(z) = \widehat J_\l(z) + \sum_{I=1}^g J_\l^I \omega_I(z),\quad
\bar J_\lb(z) = \widehat {\bar J}_\lb(z) 
+ \sum_{I=1}^g \bar J_\lb^I \omega_I(z),
$$
$$
d_\a(z) = \widehat d_\a(z) + \sum_{I=1}^g d_\a^I \omega_I(z),$$
$$
S_{mn}(z) = \widehat S_{mn}(z) + \sum_{I=1}^g S_{mn}^I \omega_I(z),\quad
S(z) = \widehat S(z) + \sum_{I=1}^g S^I \omega_I(z),$$
where $\omega_I(z)$ are the $g$ holomorphic one-forms satisfying
$\int_{A_I} dz ~\omega_J(z) = \d_{IJ}$, $\int_{A_I} dz$ are contour
integrals around the $g$ non-trivial $A$-cycles, and the hatted variables
$\widehat F(z)$ of \separate\
have no zero modes and are defined
to satisfy $\int_{A_I} dz \widehat F(z)=0$ for
$I=1$ to $g$.

As in multiloop calculations using
the minimal pure spinor formalism \loop, one can use the poles
in the OPE's of \ope\
and \opebar\ for the hatted variables
to perform the functional integral over the non-zero modes. 
Note that the partition function for the non-zero modes
is equal to one since there are an equal number
of bosons and fermions at $+1$ conformal weight.

After integrating out the non-zero modes, one obtains 
\eqn\corrfour{{\cal A} = \langle {\cal N}
f(\l,\lb,r,\t,N_{mn}^I,\bar N_{mn}^I,
J_\l^I, \bar J_\lb^I, d_\a^I, S_{mn}^I, S^I) \rangle}
where $f$ is some BRST-invariant function of the zero modes with
U(1) charge $3-3g$.
To regularize
this integral over the zero modes, the factor ${\cal N}(y)$ will be chosen as
${\cal N}(y) = \exp (\{Q,\chi(y)\})$ where
\eqn\chinew{\chi (y)
= -\lb_\a(y) \t^\a(y) - \sum_{I=1}^g (\half N_{mn}^I S^{mn I} + J_\l^I S^I).}
Using the BRST transformations
\eqn\brstt{\{Q, S_{mn}^I\}=\bar N_{mn}^I,\quad 
\{Q, S^I\}=\bar J_\lb^I,}
$$
[Q, N_{mn}^I]=-\half\int_{A_I} dz \l\g_{mn}d,\quad 
[Q, J_\l^I]=\int_{A_I} dz \l^\a d_\a,$$
one finds that 
\eqn\calNloop{{\cal N}(y) = 
\exp ( -\lb_\a(y)\l^\a(y) -r_\a (y)\t^\a(y))}
$$\exp(~
\sum_{I=1}^g [~-\half N_{mn}^I \bar N^{mn I} - J_\l^I \bar J_\lb^I ~~
- {1\over 4} S_{mn}^I
\int_{A_I} dz ~\l\g^{mn}d  ~~
+ S^I
\int_{A_I} dz ~\l^\a d_\a ~
]~).$$

So one needs to compute the integral over the zero modes
\eqn\intloop{{\cal A} = \int [d\l][d\lb][dr] d^{16}\t
\prod_{I=1}^g [dw^I] [d\bar w^I] [ds^I] d^{16} d^I
~~{\cal N} f.}
Using the methods of \loop, one can show that the measure factors
$[dw^I]$, $[d\bar w^I]$, and $[ds^I]$ are defined as
\eqn\measurew{[dw^I]\l^{\a_1} ...\l^{\a_8} = 
M_{m_1 n_1 ... m_{10} n_{10}}^{\a_1 ... \a_8}
dN^{m_1 n_1 I} 
... dN^{m_{10} n_{10} I} dJ^I_\l,}
$$[d\bar w^I]\lb_{\a_1} ...\lb_{\a_8} = 
(M^{-1})^{m_1 n_1 ... m_{10} n_{10}}_{\a_1 ... \a_8}
d\bar N^I_{m_1 n_1} 
... d\bar N^I_{m_{10} n_{10}} d\bar J^I_\lb, $$
$$[ds^I] = 
M_{m_1 n_1 ... m_{10} n_{10}}^{\a_1 ... \a_8} \lb_{\a_1} ... \lb_{\a_8}
{\p\over{\p S^I_{m_1 n_1}}}
... {\p\over{\p S^I_{m_{10} n_{10}}}} {\p\over{\p S^I}},$$
where 
$ M_{m_1 n_1 ... m_{10} n_{10}}^{\a_1 ... \a_8} $ is a Lorentz-invariant
tensor which 
is antisymmetric after switching $m_j$ with $n_j$, antisymmetric
after switching $[m_j n_j]$ with $[m_k n_k]$, and symmetric and gamma-matrix
traceless in $(\a_1 ... \a_8)$. Up to an overall normalization constant,
\eqn\overalln{
 M_{m_1 n_1 ... m_{10} n_{10}}^{\a_1 ... \a_8} \lb_{\a_1} ...\lb_{\a_8}
\psi^{m_1 n_1} ...\psi^{m_{10} n_{10}}=}
$$(\lb\g_{m_1 n_1 m_2 m_3 m_4}\lb)
(\lb\g_{m_5 n_5 n_2 m_6 m_7}\lb)
(\lb\g_{m_8 n_8 n_3 n_6 m_9}\lb)
(\lb\g_{m_{10} n_{10} n_4 n_7 n_9}\lb) 
\psi^{m_1 n_1} ...\psi^{m_{10} n_{10}} $$
where $\psi_{m_j n_j}$ are fermionic antisymmetric two-forms.

As long as $f$ does not diverge too fast as $\l\lb\to 0$, the 
regularized expression of \intloop\ is well-defined.
For 
example, if $f$ is assumed to be independent
of $S_{mn}^I$ and $S^I$, then all $11 g$ zero modes for these 
fermionic variables
must come from the regularization factor ${\cal N}$ of \calNloop.
Each of these zero modes is multiplied by a factor of $(\lambda \g_{mn} d)$
or $(\lambda^\a d_\a)$,
so ${\cal N}$ contributes a factor which goes like $\l^{11g}$ as $\l\lb\to 0$.
Since $[d\l][d\lb][dr]\to \l^8\lb^{11}$ and
$\prod_{I=1}^g [dw^I][d\bar w^I][ds^I] \to \l^{-8g}$, 
$\int [d\l][d\lb][dr]\prod_{I=1}^g [dw^I] [d\bar w^I] [ds^I]
{\cal N} $ goes like $\l^{8+3g}\lb^{11}$ as $\l\lb\to 0$.

So $f$ must diverge slower than
$\l^{-8-3g}\lb^{-11}$ as $\l\lb\to 0$ in order that \intloop\
is well-defined. Since each $b$ ghost goes like $\lb/(\lb\l)^4$
as $\l\lb\to 0$, the regularization method described here is valid
for three or fewer $b$ ghosts, i.e. for amplitudes up to two loops. 
To compute amplitudes with more than two loops using the topological
string methods described here, one needs to find an alternative
regularization method for the zero modes. Work is currently in progress
with Nikita Nekrasov on finding such a method.

To check the consistency of this computational method, consider
the zero mode structure of 
four-point massless one-loop and two-loop amplitudes.
At one-loop, there is one $b$ ghost of \bgh, one unintegrated vertex
operator $V=\l^\a A_\a(x,\t)$, and three integrated vertex operators
\eqn\intmass{U=\int dz(\p\t^\a A_\a + \Pi^m B_m + d_\a W^\a + N_{mn} F^{mn})}
where $(A_\a,B_m)$ are the spinor and vector gauge superfields and
$(W^\a,F_{mn})$ are the spinor and vector field-strengths of super-Yang-Mills.
To absorb the 16 $d_\a$ and 11 $s^\a$ fermionic zero modes, 
${\cal N}$ must contribute 11 $d_\a$ and 11 $s^\a$ zero modes,
the $b$ ghost must contribute 2 $d_\a$ zero modes through the term
$(\lb\g^{mnp}r)(d\g_{mnp}d)/(\lb\l)^2$, and each of the three integrated vertex
operators must contribute a $d_\a$ zero mode through the term $\int dz d_\a
W^\a$. After integrating over the $r_\a$ zero modes, the amplitude is
proportional to 
\eqn\onel{\int d^{16}\t (\t)^{10} A W W W,}
where the Lorentz
contractions of the spinor indices has not yet been worked out.
However, by dimensional analysis, one see that \onel\
has the correct zero mode structure to contribute an $F^4$ term
for open strings, or an $R^4$ term for closed strings after taking
the holomorphic square. 

For four-point two-loop massless amplitudes, 
there are three $b$ ghosts of \bgh\
and four integrated vertex operators
of \intmass. To absorb the 32 $d_\a$ and 22 $s^\a$ fermionic zero modes,
${\cal N}$ must contribute 22 $d_\a$ and 22 $s^\a$ zero modes, each of
the three $b$ ghosts must contribute
2 $d_\a$ zero modes through the term
$(\lb\g^{mnp}r)(d\g_{mnp}d)/(\lb\l)^2$, and each of
the four integrated vertex
operators must contribute a $d_\a$ zero mode through the term $\int dz d_\a
W^\a$. After integrating over the $r_\a$ zero modes, the amplitude is
proportional to 
\eqn\twol{\int d^{16}\t (\t)^{8} W W W W,}
which has the correct zero mode structure to contribute a $\p^2 F^4$ term
for open strings, or a $\p^4 R^4$ term for closed strings after taking
the holomorphic square.
It should not be too difficult to verify if 
the contractions of the Lorentz indices in \onel\ and \twol\ reproduce
the appropriate $t_8$ index contractions in the $R^4$ and $\p^4 R^4$ terms.

\newsec{Cubic Open Superstring Field Theory}

Using the RNS formalism for the superstring, cubic open superstring
field theory actions require midpoint insertions which cause 
contact-term divergences or gauge invariance problems. For example,
in the cubic Neveu-Schwarz action of \ref\witsuper{E. Witten,
{\it Interacting field theory of open superstrings}, Nucl. Phys.
B276 (1986) 291.},
\eqn\cubone{S = \langle \half V Q V + {1\over 3} V V V Z({\pi\over 2})
\rangle,} 
where the open string fields $V$ are multiplied using Witten's star product,
$V$ is chosen in the $-1$ picture, and $Z({\pi\over 2})$ is
the picture-raising operator inserted at the string midpoint.
Since $Z(y) Z(z)$ is divergent when $y\to z$, the action produces
unphysical contact-term divergences when interaction points collide \ref\wendt
{C. Wendt, {\it Scattering amplitudes and contact interactions in
Witten's superstring field theory}, Nucl. Phys. B314 (1989) 209.}\ref
\greensite{J. Greensite and F.R. Klinkhamer, {\it New interactions for
superstrings}, Nucl. Phys. B281 (1987) 269.}.
Alternatively, in the cubic Neveu-Schwarz action of \ref\thorn
{C.R. Preitschopf, C.B. Thorn and S.A. Yost, {\it Superstring field theory},
Nucl. Phys. B337 (1990) 363.}\ref\aref{I.Ya. Arefeva, D.M. Belov, 
A.S. Koshelev and P.B. Medvedev, {\it Tachyon condensation in cubic
superstring field theory}, Nucl. Phys. B638 (2002) 3, hep-th/0011117.},
\eqn\cubtwo{S = \langle (\half V Q V + {1\over 3} V V V )
Y^2({\pi\over 2})
\rangle,} 
where $V$ is chosen in the zero picture and $Y^2({\pi\over 2})$ is
the square of the picture-lowering operator inserted at the string midpoint.
Although the action of \cubtwo\ does
not have contact-term divergences, it has gauge invariance problems since
the linearized equation of motion is $Y^2({\pi\over 2})QV=0$ instead of
$QV=0$. Since $Y^2({\pi\over 2})$ has a non-trivial kernel, the equation
$Y^2({\pi\over 2})QV=0$ has additional solutions given by 
$V=Ker(Y^2({\pi\over 2}))$.
If one projects out states in the kernel of $Y^2({\pi\over 2})$ to
remove these unwanted solutions from the Hilbert space,
the associativity property of the star-product is ruined and 
gauge invariance is broken \ref\areftwo{I.Ya. Arefeva and P.B. Medvedev,
{\it Anomalies in Witten's field theory of the NSR string},
Phys. Lett. B212 (1988) 299\semi
I.Ya. Arefeva and P.B. Medvedev,
{\it Truncation, picture-changing operation and spacetime supersymmetry
in Neveu-Schwarz-Ramond string field theory}, Phys. Lett. B202 (1988) 510.}
\ref\berkrev{N. Berkovits, {\it
Review of open superstring field theory}, hep-th/0105230.}.

Although these 
problems are avoided in the non-polynomial WZW-like action
for open superstring field theory \ref\osft{N. Berkovits,
{\it Super-Poincar\'e invariant superstring field theory},
Nucl. Phys. B450 (1995) 90, hep-th/9503099.} which does not require midpoint
insertions, it would be useful to have a cubic open superstring
field theory action.
Since
the equation of motion in the pure spinor formalism for the open superstring
field $V$ is 
\eqn\equa{QV + V V=0,}
a natural suggestion \schw\ is to use
the Chern-Simons-like action 
\eqn\action{S = \langle \half V Q V + {1\over 3} V V V \rangle}
of bosonic string field theory.
However, using the minimal pure spinor formalism of \pureone,
the inner product for zero modes defined by
\eqn\normone{\langle 0| 
(\l\g_m\t)(\l\g_n\t)(\l\g_p\t)(\t\g^{mnp}\t) |0\rangle =1}
is degenerate, so the action of \action\ does not generate the equations
of \equa.
Since the norm is degenerate, $\langle A | B \rangle=0$ for every string
field $|B\rangle$ does not imply that $| A\rangle =0$. 
For example, $|A\rangle = (\theta)^n |0\rangle$ for $n>5$ satisfies
$\langle A | B \rangle =0$ for any string
field $|B\rangle$.
Therefore, using the minimal inner product of \normone, the action of \action\
does not imply that components of  $(QV + V V)$
with more than five $\t$'s must vanish on-shell.

As shown in \loop, the inner product for zero modes
in the minimal pure spinor formalism
can be made non-degenerate by defining 
\eqn\normtwo{\langle 0| f(\l,\t) |0\rangle = \int [d\l] d^{16}\t f(\l,\t)}
where $[d\l]$ is defined in \measurelambda. This implies that
$$\langle 0|(\l\g_m\t)(\l\g_n\t)(\l\g_p\t)(\t\g^{mnp}\t) 
 \prod_{I=1}^{11} Y_{C_I} |0\rangle$$
is non-zero where $Y_{C_I}= (C^I_\a\t^\a)\d(C^I_\b\l^\b)$ is the
picture-lowering operator and $C^I_\a$ are constant spinors for $I=1$
to 11. Using this non-degenerate norm, the appropriate open superstring
field theory action would be 
\eqn\actiontwo{S = \langle (\half V Q V + {1\over 3} V V V)
\prod_{I=1}^{11} Y_{C_I}({\pi\over 2}) \rangle,}
where the eleven picture-lowering operators are inserted at the
string midpoint. However, in addition to causing gauge-invariance
problems as in the RNS cubic action of \cubtwo, 
these midpoint insertions break
Lorentz invariance because of their explicit dependence on $C_\a^I$.

As discussed in the previous section, the non-minimal pure spinor
formalism does not require picture-changing operators but instead
introduces the regularization
factor ${\cal N}=\exp(-\l\lb-r\t)$.
Since
the inner product for zero modes defined by
\eqn\normthree{\langle 0| {\cal N} f(\l,\lb,r,\t) |0\rangle = 
\int [d\l][d\lb][dr] d^{16}\t f(\l,\lb,r,\t) \exp(-\lb_\a\l^\a -r_\a\t^\a)}
is non-degenerate,
the cubic action 
\eqn\cubthree{S= \langle (\half V Q V + {1\over 3} V V V) 
{\cal N}({\pi\over 2})\rangle }
generates the equation of motion
\eqn\eqthree{ 
{\cal N}({\pi\over 2})(Q V + V V)=0,}
where the regularization factor ${\cal N}(y)$ is inserted at the
string midpoint.
But unlike the picture-lowering operator in \cubtwo\ or \actiontwo,
${\cal N}$ has no kernel
since ${\cal N}^{-1} = \exp (\l\lb + r\t)$ is well-defined even when
acting on
off-shell states. So there are no gauge invariance problems and \eqthree\
implies the desired equation of motion $QV + V V=0$.

Note that the action of \cubthree\ is manifestly Lorentz invariant,
but is not manifestly spacetime supersymmmetric because of the explicit
$\t$ dependence in the regularization factor
${\cal N} = \exp(-\lb_\a\l^\a -r_\a\t^\a)$.
The action differs from the ``minimal'' cubic action of \action\
since the string field $V$ can depend on the non-minimal variables
$\lb_\a$ and $r_\a$. Although the linearized on-shell string field
is independent of these non-minimal variables, the off-shell dependence
on the non-minimal variables is necessary for generating the $(\t)^n$
components for $n>5$ of the equation of motion $QV+V V=0$.

Although the discussion of the inner product has focused up to now
on the zero mode dependence of the string field $V$, it is easy to
see that the non-zero modes do not cause any problems. To evaluate
the cubic action of \cubthree\ for an arbitrary string field $V$,
first convert the string field to a vertex operator on the disk, and
then use the conformal field theory OPE's of \ope\ and \opebar\ for
the variables of $+1$ conformal weight to functionally
integrate over the non-zero modes. The remaining dependence on the zero
modes is integrated using the regularization factor ${\cal N}=
\exp(-\lb_\a\l^\a-r_\a\t^\a)$ as in \normthree. Since the string
field $V$ will be required to be non-singular as $\l\lb\to 0$, the integral
$\int [d\l][d\lb][dr]d^{16}\t ~{\cal N} f(\l,\lb,r,\t)$
is guaranteed to be well-defined.

For BRST-invariant external states, rescaling the regularization factor as
$${\cal N} = \exp(-\lb\l-r\t) \to {\cal N}_\rho = \exp (-\rho(\lb\l+r\t)) $$
for any positive $\rho$ does not affect the scattering amplitude.
However, since the string field $V$ is off-shell, the cubic open
superstring field theory action will depend on the scaling factor
$\rho$. To make this
dependence explicit, define the BRST-invariant charge 
\eqn\jlb{j_\lb = \int dz \bar J_\lb = 
\int dz (\bar w^\a \lb_\a - s^\a r_\a)}
such that $\lb_\a$ and $r_\a$ carry $+1$ charge and $\bar w_\a$ and $s^\a$
carry $-1$ charge. Since $\bar w^\a$ and $s^\a$ can only appear in the
$j_\lb$-neutral combinations of \gaugeinvbar\ and \other,
all states in the Hilbert space carry non-negative 
$j_\lb$ charge. And since $[Q, j_\lb]=0$, 
the cubic action of \cubthree\ can be written as $S(\rho)
= \sum_{m=0}^\infty S_m(\rho)$
where
\eqn\actionm{S_m(\rho) = \langle (\half \sum_{p=0}^m  V_p Q V_{m-p}
+ {1\over 3} \sum_{p=0}^m\sum_{q=0}^{m-p} V_p V_q V_{m-p-q} )
{\cal N}_\rho({\pi\over 2})\rangle}
and $V_q$ is a string field satisfying $j_\lb (V_q) = q V_q$.
Under the scaling 
of $\lb_\a \to c\lb_\a$ and $r_\a\to c r_\a$, 
one can easily verify that ${\cal N}_\rho \to {\cal N}_{c\rho}$,
$V_q \to c^q V_q$, 
and the measure factor $[d\lb][dr]$ is invariant. 
This implies that $S_m(\rho)= \rho^{-m} S_m(1)$ 
and that the dependence of $S$ on $\rho$ can be cancelled by
rescaling the string field as $V_q\to \rho^q V_q$.
Note that all propagating on-shell string fields have zero $j_\lb$ charge,
so they are unaffected by the rescaling of the regularization factor.

For closed topological strings describing Calabi-Yau three-folds, it
is possible to construct a cubic closed string field theory action which
resembles the action for Kodaira-Spencer gravity \kodaira. 
It would be very interesting
to see if this construction for closed topological strings generalizes
to the non-minimal pure spinor formalism for closed superstring field
theory. Since the closed string field
theory action involves the $b$ ghost, this generalization may not be
straightforward because of the singularitites in the $b$ ghost of \bgh\ when
$\l\lb\to 0$. However, it is encouraging that the kinetic term for the
Ramond-Ramond sector of closed superstring field theory \ref\emdu
{N. Berkovits, {\it Manifest electromagnetic duality in closed
superstring field theory}, Phys. Lett. B388 (1996) 743, hep-th/9607070.}
can be constructed
using a set of non-minimal variables which have some similarities
with the non-minimal variables of the pure spinor formalism.

\newsec{Four-dimensional Pure Spinor Formalism}

\subsec{Minimal $d=4$ pure spinor formalism}

Since topological strings are useful for computing superpotential terms
in the four-dimensional spacetime action \antoniadis\kodaira, 
it is natural to look for
a four-dimensional version of the pure spinor formalism. In four dimensions,
the Green-Schwarz-Siegel matter variables consist of $(x^m,\t^a,\tb^\ad,p_a,
\bar p_\ad)$ for $m=0$ to 3 and $a,\dot a =1$ to 2, where $p_a$ and 
$\bar p_\ad$
are the conjugate momenta for $\t^a$ and $\tb^\ad$.
Since a $d=4$ pure spinor is simply a chiral two-component spinor $\l^a$,
the natural $d=4$ version of the ``minimal'' pure spinor formalism is
constructed from
the $d=4$ Green-Schwarz-Siegel variables, a $\hat c=3$ N=2 superconformal field
theory for the six-dimensional compactification manifold, and a $d=4$
pure spinor ghost $\l^a$ together with its conjugate momentum $w_a$.
The worldsheet action for these variables is
\eqn\actionfour{S = \int d^2 z (\half \p x^m \bar\p x_m + p_a\bar\p\t^a
+\bar p_\ad\bar\p\tb^\ad - w_a\bar\p\l^a) + S_C}
where $S_C$ is the worldsheet action for the compactification-dependent
variables. 

The worldsheet variables in \actionfour\
are the same as in the $d=4$ hybrid formalism \hybrid\
for the superstring
except for the replacement of $(\l^a,w_a)$
with a chiral boson $\rho$ satisfying the OPE $\rho(y)\rho(z)\to -\log(y-z)$. 
Recall that in the $d=4$ hybrid formalism, physical states are defined as
N=2 primary fields with respect to the $\hat c=2$ N=2 generators
\eqn\hyb{J = -\p\rho + J_C,\quad G^+ = e^\rho d_a d^a + G_C^+,\quad
G^- = e^{-\rho} \bar d_\ad \bar d^\ad + G_C^-,}
$$
T = 
-\half \p x^m \p x_m - p_a\p\t^a
-\bar p_\ad\p\tb^\ad -\half \p\rho\p\rho + T_C$$
$$=-\half \Pi^m \Pi_m - d_a\p\t^a
-\bar d_\ad\p\tb^\ad -\half \p\rho\p\rho + T_C,$$
where 
$d_a = p_a +{i\over 2} \p x_m \s^m_{a\ad}\tb^\ad 
-{1\over 4}(\tb)^2\p\t_a +{1\over 8}\t_a \p(\tb)^2,$
$\bar d_\ad = \bar p_\ad +{i\over 2} \p x_m \s^m_{a\ad}\t^a 
-{1\over 4}(\t)^2\p\tb_\ad +{1\over 8}\tb_\ad \p(\t)^2$,
$\Pi^m = \p x^m -{i\over 2}\s^m_{a\ad}(\tb^\ad \p\t^a +\t^a\p\tb^\ad)$, and
$[J_C, G^+_C, G^-_C, T_C]$ are the $\hat c=3$ N=2 superconformal
generators for the compactification manifold.
After twisting, the N=2 generators of \hyb\ are related by a field
redefinition to the RNS operators
\eqn\rns{J = bc + \eta\xi,\quad G^+ = j_{BRST}^{RNS}\quad
G^- = b, \quad
T = T^{RNS}_{matter} + T^{RNS}_{ghost},}
and the N=2 physical state condition is mapped to the usual requirement of
BRST-invariance for RNS physical states.

In the ``minimal'' version of the $d=4$ pure spinor formalism, physical
states will instead be defined as ghost-number one states in the cohomology
of the ``minimal'' BRST operator
\eqn\fourmin{Q = \int dz (\l^a d_a + G_C^+)}
where the ghost-number is defined by the charge
\eqn\jghost{j_{ghost} = \int dz (w_a\l^a + J_C)}
and $[J_C, G_C^+, G_C^-, T_C]$ are the twisted $\hat c=3$ N=2
superconformal generators for the compactification manifold.
To compute the cohomology of $Q$, it is convenient to perform
a similarity transformation on the worldsheet variables so that
\eqn\chiralsuper{d_a = p_a,\quad
\bar d_\ad = \bar p_\ad + i\p x_m \s^m_{a\ad}
\t^a -  (\t)^2 \p\tb_\ad,
\quad \Pi^m =\p x^m - i\t^a\s^m_{a\ad}\p\tb^\ad,}
as in a chiral $d=4$ superspace representation.
Since states in the cohomology of $\int dz(\l^a p_a)$ are independent
of $(\t^a,p_a,\l^a, w_a)$, any ghost-number one
state in the cohomology of $Q$ can
be expressed as
\eqn\solcoh{V = \Phi_j(x,\tb,\bar p)\psi^j}
where $\Phi_j$ is a superfield depending on both zero modes
and non-zero modes of $(x^m,\tb^\ad,\bar p_\ad)$, and $\psi^j$ is a chiral
primary of $+1$ charge with respect to the $\hat c=3$
N=2 superconformal field theory for the compactification manifold.

Since $\Phi_j$ can depend on the non-zero modes of $(x^m,\tb^\ad,\bar p_\ad)$,
$V$ describes both massive and massless states, and the $d=4$ mass-shell
condition is not imposed by BRST invariance. As will now be explained,
$V$ describes the chiral sector of open superstring field theory which
contributes to F-terms in the open superstring field theory action. 
So the $d=4$ pure spinor formalism can be understood as a $d=4$
super-Poincar\'e covariant version of the $\hat c=5$ topological
string of \oog.

When written in terms of $d=4$ superspace variables using the hybrid formalism,
the open superstring field theory action \osft\berkrev\ depends on three 
string fields which contain $J_C$ charge $+1$, $0$, and $-1$.
The string field with zero $J_C$ charge describes 
compactification-independent fields like the 
$N=1$ $d=4$ super-Yang-Mills multiplet, the string field with $+1$
$J_C$ charge describes compactification-dependent fields like the
chiral moduli, and the string field with $-1$
$J_C$ charge describes compactification-dependent fields like the
anti-chiral moduli. Although the D-term in the open superstring field
theory action contains couplings between all three string fields,
the F-term only involves the string field with $+1$ $J_C$ charge
which will be called $V$.

Using the language of the $d=4$ hybrid formalism, $V$ is restricted
to satisfy $[\int dz G_4^+, V]=0$ where $G_4^+ = e^\rho d_a d^a$, which implies
that $V$ has no poles with $d_a$. The F-term in the open superstring
field theory action is given by \osft\berkrev
\eqn\fone{S = \langle \half V (\int dz G_C^+) V + {1\over 3} V V V\rangle_F}
where $G_C^+$ is the spin-one fermionic generator from the twisted
$\hat c=3$ N=2 superconformal field theory for the compactification
manifold, $\langle ~~~\rangle_F$ denotes the norm for F-terms defined
by $\langle J^{+++}_C (\tb)^2\rangle =1$, and
$J^{+++}_C$ is the spectral-flow operator with $+3$ $J_C$ charge
for the $\hat c=3$ N=2 superconformal field theory that
describes the compactification
manifold.

To understand the definition of $\langle ~~~\rangle_F$, note that the
norm $\langle ~~~\rangle_D$ for D-terms is defined by
$\langle J_C^{+++} e^{-\rho} (\t)^2 (\tb)^2 \rangle_D=1$, which maps
to $\langle c \p c\p^2 c\xi e^{-2\phi}\rangle_D =1$ using the field
redefinition to the RNS formalism. Since
\eqn\relgen{[\int dz G_4^+, J_C^{+++} e^{-\rho}
(\t)^2(\tb)^2] = J_C^{+++}(\tb)^2,}
one finds
that $\langle [\int dz G_4^+, A]\rangle_F = \langle A\rangle_D$ for
any function $A$, which is the superstring generalization of the
usual superspace relation between
F-terms and D-terms that
$\langle D_a D^a A\rangle_F = \langle A\rangle_D$ where $D_a$ 
are the N=1 $d=4$ chiral superspace derivatives.

For example, for compactification on $T^6$ where the worldsheet variables
are $(y^j,\psi^j)$ and $(\bar y_j,\bar\psi_j)$ for $j=1$ to 3, the
twisted $\hat c=3$ N=2 generators for the compactification manifold are
$T_C = \p y^j \p \bar y_j + \bar\psi_j \p \psi^j$, 
$G_C^+ =\p \bar y_j \psi^j$,
$G_C^- = \p y^j \bar\psi_j$ and $J_C = \psi^j \bar\psi_j$. Besides
depending on chiral superfields coming from
Kaluza-Klein reduction of $d=10$ massive multiplets,
the string 
field $V$ depends on three chiral superfields 
$\Sigma_j(x,\tb,y,\bar y)$ which come from Kaluza-Klein reduction of the $d=10$
massless super-Yang-Mills multiplet. The dependence of the string field on
these superfields is given by
$V = \Sigma_j(x,\tb,y,\bar y)\psi^j$, and after
plugging $V$ into \fone, one obtains
the expected F-term
\eqn\massr{S=\int d^4 x \int d^6 y \int d^2 \tb
\epsilon^{ijk}
Tr (\half \Sigma_i\p_j\Sigma_k + {1\over 3}\Sigma_i\Sigma_j\Sigma_k)}
for these superfields \ref\siegelma{N. Marcus, A. Sagnotti and
W. Siegel, {\it Ten-dimensional supersymmetric Yang-Mills theory in
terms of four-dimensional superfields}, Nucl. Phys. B224 (1983) 159.}.

If one drops the D-term in the open superstring field theory action and
keeps only the F-term of \fone, physical states are described by a string
field $V$ with $+1$ $J_C$ charge, with no poles with $d_a$, and
which satisfies the linearized equation of motion $\{\int dz G_C^+ ,V\}=0$
with the linearized gauge invariance $\d V = [\int dz G_C^+, \Omega]$.
So physical states defined with respect to \fone\
carry $+1$ $J_C$ charge, are independent of $\t^a$, and are chiral primaries
with respect to the $\hat c=3$ N=2 superconformal field theory for
the compactification manifold. Since this definition of physical states
coincides with the definition of physical states in the $d=4$ pure
spinor formalism, it is natural to conjecture that the 
$d=4$ pure spinor formalism describes the chiral sector of superstring
theory which contributes to F-terms in the superstring field theory action.
Evidence for this conjecture will now be provided by computing
scattering amplitudes using the non-minimal version of the $d=4$ pure spinor
formalism.

\subsec{Non-minimal $d=4$ pure spinor formalism}

In analogy with the non-minimal version of the $d=10$ pure spinor formalism,
the $d=4$ non-minimal variables will consist of a bosonic chiral spinor
$\lb_a$ and fermionic chiral spinor $r_a$, with conjugate momentum
$\bar w^a$ and $s^a$. 
Since chiral two-component spinors are automatically $d=4$ pure spinors,
there are no additional constraints on $\lb_a$ and $r_a$ analogous to
the $d=10$ constraints of \newcons.
Although it might seem strange that the $d=4$ non-minimal
variables have the same spacetime chirality as $\l_a$ whereas the
$d=10$ non-minimal variables had the opposite spacetime chirality, note that in
four dimensions, complex conjugation in Euclidean space does not flip the
chirality of spacetime spinors. 
The worldsheet action including the non-minimal variables is
\eqn\actionfour{S = \int d^2 z (\half \p x^m \bar\p x_m + p_a\bar\p\t^a
+\bar p_\ad\bar\p\tb^\ad - w_a\bar\p\l^a - \bar w^a \bar\p\lb_a 
+s^a\bar\p r_a) + S_C,}
where the barred $(\tb^\ad,\bar p_\ad)$ variables 
will be defined to carry dotted spinor indices
while the barred $(\lb_a,\bar w^a)$ variables will
carry undotted spinor indices.

In order that the non-minimal variables do not affect the cohomology,
the ``minimal'' pure spinor BRST operator $Q=\int dz (\l^a d_a + G_C^+)$ will
be modified to the ``non-minimal'' BRST operator
\eqn\nonminQf{Q = \int dz (\l^a d_a + \bar w^a r_a + G_C^+).}
It is straightforward to construct a $b$ ghost satisfying
$\{Q, b\}= T$  and one finds
\eqn\fbgh{b =  s^a\p\lb_a + w_a\p\t^a  + G_C^- +
{i{\lb_a 
\Pi^m \bar\s_m^{\ad a}\bar d_\ad }\over{2(\lb\l)}} 
- {{(\e^{ab}\lb_a r_b)(\e^{\ad\bd}\bar d_\ad \bar d_\bd)}
\over{4(\lb\l)^2}} ,}
where
\eqn\Tf{T= -\half \p x^m \p x_m - p_a \p\t^a -\bar p_\ad\p\bar\t^\ad
 + w_a \p\l^a
+ \bar w^a \p\lb_a - s^a \p r_a +T_C}
$$= -\half \Pi^m \Pi_m - d_a \p\t^a -\bar d_\ad\p\bar\t^\ad
 + w_a \p\l^a
+ \bar w^a \p\lb_a - s^a \p r_a +T_C,$$
and $d_a$, $\bar d_\ad$ and $\Pi^m$ are defined in \chiralsuper.

One can verify that $b$ has no poles with itself and that the double pole
of $b$ with the BRST integrand $j_{BRST} = \l^a d_a + \bar w^a r_a + G_C^+$
produces the U(1) generator
\eqn\Jfour{J = \l^a w_a + r_a s^a + J_C.}
The generators $[J, j_{BRST}, b, T]$ form a $\hat c=3$ N=2 algebra
which allow the formalism to be interpreted as a critical topological string.
However, as in the $d=10$ non-minimal pure spinor formalism, it is
convenient to shift the U(1) generator by a BRST-trivial quantity
$\{Q, - s^a \lb_a\} = -\bar w^a \lb_a -r_a s^a$ so that the new ghost
charge is
\eqn\newfourc{j_{ghost} =\int dz J = \int dz (\l^a w_a + r_a s^a + J_C
+ \{Q, -s^a \lb_a\}) = \int dz (\l^a w_a - \lb_a \bar w^a + J_C).}

The standard topological rules for computing scattering amplitudes
can now be applied using
the BRST operator of 
\nonminQf, the $b$ ghost of \fbgh, the stress tensor of \Tf,
and the ghost charge of \newfourc.
For example,
$N$-point tree amplitudes are computed by the correlation function
\eqn\fcorrtree{{\cal A} = \langle {\cal N}(y) V_1(z_1) V_2(z_2) V_3(z_3)
\int dz_4 U_4(z_4) ... \int dz_N U_N(z_N) \rangle}
where, 
as in \corrtree,
the regularization factor 
\eqn\fconv
{{\cal N} = \exp(\{Q,-\lb_a\t^a\}) = \exp (-\lb_a\l^a -r_a\t^a)}
will be inserted into the correlation function.

After integrating out the worldsheet non-zero modes,
the zero mode integral is
\eqn\fzm{\langle
{\cal N}~ f(\l,\lb,r,\t,\tb,\psi)\rangle
=
\int d^2 \l d^2 \lb d^2 r d^2\t d^2\tb d^3 \psi 
\exp (-\lb_a\l^a -r_a\t^a) f(\l,\lb,r,\t,\tb,\psi),}
which
is well-defined as long as
$f(\l,\lb,r,\t,\tb,\psi)$ diverges slower than
$(\l\lb)^{-2}$ as $\l\lb\to 0$.

The restriction that $f(\l,\lb,r,\t,\tb,\psi)$ 
diverges slower than $(\l\lb)^{-2}$
is related to the operator $\xi = (\lb\t)/(\lb\l + r\t)$ which satisfies
$Q\xi=1$. Using the same argument as in the $d=10$ non-minimal pure
spinor formalism,
$\langle {\cal N}~Q\Omega\rangle$ is only guaranteed to
vanish if $\Omega$ diverges slower than $\l^{-2}\lb^{-1}$ as $\l\lb\to 0$.
This allows $\langle {\cal N}~ f\rangle$ to be 
non-vanishing when $f=(\tb)^2 (\psi)^3$
since although $\langle {\cal N}~ f\rangle = \langle {\cal N}~
Q(\xi f)\rangle$, $\xi f$
diverges like
$(\t)^2(\tb)^2 (\psi)^3 (\lb r)/(\l\lb)^2$ when $\l\lb\to 0$. 

Returning to the $N$-point tree amplitude computation, suppose that
all external states are chosen in the gauge where the vertex operators
are independent of the non-minimal fields. Then after integrating out
the non-zero modes, one obtains
\eqn\fintthree{{\cal A} = \int d^2\l d^2\lb d^2 r d^2 \t d^2 \tb d^3 \psi
\exp (-\lb_a\l^a -r_a\t^a) (\psi)^3 f(\tb),}
where all ghost charge in the vertex operators must come from the 
compactification-dependent
variables $\psi^j$ since states in the cohomology are independent of $\l^a$
and $\t^a$. Integrating over $\l^a$, $\lb_a$ and $r_a$, one finds
\eqn\fintfour{{\cal A} = \int d^2 \t d^2 \tb d^3 \psi
~~(\t)^2 (\psi)^3 f(\tb) = \int d^2\tb f(\tb),}
which is the desired result for the F-term in the scattering amplitude.

One can also compute $N$-point tree amplitudes in a worldsheet
reparameterization invariant manner using $(N-3)$ $b$ ghosts and
$N$ integrated vertex operators as 
\eqn\corrtwo{{\cal A} = \langle {\cal N}(y) V_1(z_1) ... V_N(z_N)
\int dz_4 b(z_4) ... \int dz_N b(z_N) \rangle}
where $b(z)$ is defined in \fbgh\ and ${\cal N}$ is defined in
\fconv. But unlike the $d=10$ computation, there
is no restriction on the number of $b$ ghosts 
in the $d=4$ computation. This is because all ghost charge in
physical states must come from the compactification-dependent
variables, so each 
unintegrated vertex operator contributes $+1$ $J_C$ charge.
By charge conservation of $J_C$, this implies that the only term
which contributes in $b(z)$ is the $G_C^-$ term which carries $-1$
$J_C$ charge. Since $G_C^-$ has no singularities when $\l\lb\to 0$,
there is no restriction on the number of $b$ ghosts in the $d=4$
pure spinor formalism.

To compute $N$-point $g$-loop amplitudes, one uses the 
topological prescription
\eqn\corrthree{{\cal A} = \int d^{3g-3}\tau \langle {\cal N}(y) 
\prod_{j=1}^{3g-3}(\int dw_j \mu_j(w_j) b(w_j)) \prod_{r=1}^N 
\int dz_r U(z_r) \rangle}
where $\tau_j$ are the complex Teichmuller parameters and $\mu_j$ are
the associated Beltrami differentials, 
\eqn\fcalNloop{{\cal N}(y) = 
\exp (\{Q,\chi(y)\}) = 
\exp ( -\lb_a(y)\l^a(y) -r_a(y) \t^a (y)-
\sum_{I=1}^g (w_a^I \bar w^{a I}- d_a^I s^{a I})),}
$\chi(y) 
= -\lb_a(y) \t^a(y) - \sum_{I=1}^g w_a^I s^{a I} $, and
$(w_a^I,\bar w^{aI}, s^{aI},d_a^I, \bar d_\ad^I, \bar\psi_j^I)$ for
$I=1$ to $g$
are the zero modes for the variables of $+1$ conformal weight.

After separating off the zero modes of $(w_a,\bar w^a,s^a,
d_a,\bar d_\ad,\bar \psi_j)$ as in \separate\ and integrating over
the non-zero modes, one obtains
\eqn\fcorrfive{{\cal A} = 
\int d^2\l d^2\lb d^2 r d^2\t d^2 \tb
d^3 \psi \prod_{I=1}^g
d^2 w^I d^2 \bar w^I d^2 s^I d^2 d^I d^2 \bar d^I d^3\bar\psi^I}
$$ 
{\cal N} f(\l,\lb,r,\t,\tb,\psi,w^I,\bar w^I,
d^I, \bar d^I, s^I, \bar\psi^I) $$
where
$f$ is some BRST-invariant function of the zero modes with
U(1) charge $3-3g$.
Since conservation of $J_C$ charge implies that
only the $G_C^-$ term in the $b$ ghost contributes to the $g$-loop
amplitude, $f$ has no singularities when $\l\lb\to 0$ and
there is no restriction on the number of $b$ ghosts
or on the genus $g$.

One can easily check that this prescription for the closed superstring
reproduces the $g$-loop scattering amplitude of $2g$ self-dual graviphotons
and an arbitrary number of chiral superfields for the Calabi-Yau
moduli \antoniadis\kodaira. 
As in the computation using the $d=4$ hybrid formalism \topovafa\
or using the $\hat c=5$ topological formalism \oog,
the $2g$ zero modes for $\bar d_\ad$ come from the graviphoton vertex
operators, the $3g-3$ zero modes for $\bar \psi_j$ come from the
$b$ ghosts, and the two $\bar\theta^\ad$ zero modes come from
the Calabi-Yau chiral superfields.
The remaining $2g$ zero modes for $d_a$, $2g$ zero modes for $s^a$,
two zero modes for $r_a$, and two zero modes for $\t^a$ come from
the regularization factor ${\cal N}$ of \fcalNloop.

So the topological string prescription for scattering amplitudes
using the $d=4$ pure spinor
formalism correctly reproduces the F-term in the spacetime action.
Further confirmation that the $d=4$ pure spinor formalism describes
F-terms comes from the open string field theory action for the
$d=4$ pure spinor formalism. Using the construction of section 4, 
the open string field theory action for the $d=4$ pure
spinor formalism is 
\eqn\factionsft{S = \langle (\half V Q V + {1\over 3}V V V){\cal N}
({\pi\over 2})\rangle}
where $Q$ is defined in \nonminQf\ and ${\cal N}$ is defined in
\fconv.
The action of
\factionsft\ has the same Chern-Simons structure as the F-term of
\fone\ in the open superstring field theory action, and it should
not be difficult to prove their equivalence.
It would be interesting to generalize this construction of the F-term
in non-trivial closed string backgrounds involving Ramond-Ramond fields.

\newsec{Appendix: U(5)-Covariant Variables for the Non-Minimal Formalism}

In this appendix, the constraints of \puredef\ and
\newcons\ for the pure spinor ghost
and non-minimal variables will be solved in a U(5)-covariant manner
in terms of free fields. The coefficients in the OPE's of \ope\
and \opebar\ can then
be computed using the free field OPE's of the U(5)-covariant variables.

As shown in \pureone, the pure spinor constraint $\l\g^m\l=0$ can
be solved in terms of free fields as 
\eqn\lamsol{\l^\a = (\l^+, \l_{ab},\l^a) = (\g,\g u_{ab}, -{1\over 8}\g
\e^{abcde}u_{bc}u_{de}),}
where $a=1$ to 5, $u_{ab}=-u_{ba}$, and $(\l^+,\l_{ab},\l^a)$ describe
the $(1,10,\bar 5)$ components of $\l^\a$ under the $U(5)$ decomposition of the
(Wick-rotated) SO(10) pure spinor.
In terms of the variables $(\g,u_{ab})$ and their conjugate momenta
$(\b,v^{ab})$, the gauge-invariant currents of \currone\ are 
\eqn\gif{N^{ab} = v^{ab},}
$$N_a^b = -u_{ac} v^{bc} +\d_a^b ({5\over 4}\eta\xi +{3\over 4}\p\phi),$$
$$N_{ab} = 3\p u_{ab} + u_{ac} u_{bd} v^{cd} - u_{ab} ({5\over 2}\eta\xi
+{3\over 2}\p\phi),$$
$$J_\l = -{5\over 2}\p\phi -{3\over 2}\eta\xi,$$ 
$$T_\l= \half v^{ab} \p u_{ab} -\eta\p\xi -\half(\p\phi\p\phi + \p^2\phi)
-{7\over 2}\p(\eta\xi+\p\phi),$$
where $\g=\eta e^\phi$ and $\b=\p\xi e^{-\phi}$. It is straightforward to
use the free field OPE's
\eqn\freeone{v^{ab}(y) u_{cd}(z) \to \d_c^{[a}\d_d^{b]} (y-z)^{-1},\quad
\eta(y)\xi(z)\to (y-z)^{-1},\quad \phi(y)\phi(z) \to -\log(y-z),}
to show that these currents satisfy the OPE's of \ope.

To describe the non-minimal variables $(\lb_\a,r_\a)$ in terms of
unconstrained U(5)-covariant variables, define
\eqn\lambarsol{\lb_\a = (\lb_+, \lb^{ab},\lb_a) = \bar\g (1, \bar u^{ab}, 
-{1\over 8} \e_{abcde}\bar u^{bc}\bar u^{de}),}
$$r_\a = (r_+, r^{ab},r_a) = \bar\g (f, f^{ab} +f u^{ab}, 
-{1\over 8} \e_{abcde}
(f \bar u^{bc}\bar u^{de} +2 f^{bc}\bar u^{de})),$$
which satisfy the constraints $\lb\g^m\lb = r\g^m\lb=0$ of \newcons.
In terms of the variables $(\bar\g,\bar u^{ab}, f,f^{ab})$ 
and their conjugate momenta
$(\bar\b,\bar v_{ab},g, g_{ab})$, the gauge-invariant currents of 
\gaugeinvbar\ and \other\ are 
\eqn\gifbar{\bar N_{ab} = \bar v_{ab},}
$$\bar N_a^b = \bar u^{bc} \bar v_{ac} +f^{bc} g_{ac}
+\d_a^b (
-{1\over 4}\bar\eta\bar\xi +{1\over 4}\p\bar\phi),$$
$$\bar N^{ab} = \bar u^{ac} \bar u^{bd} \bar v_{cd} + 
\bar u^{ac} f^{bd} g_{cd} + f^{ac} \bar u^{bd} g_{cd} - f^{ab} g
-\bar u^{ab} (\half\bar\eta\bar\xi
-\half\p\bar\phi),$$
$$\bar J_\lb = -\half\p\bar\phi +\half\bar\eta\bar\xi, \quad
J_r = fg +\half f^{ab} g_{ab} +8(\bar\eta\bar\xi +\p\bar\phi),\quad
\Phi = \half f^{ab}\bar v_{ab} + \half f(\bar\eta\bar\xi -\p\bar\phi),$$
$$T_\lb= \half \bar v_{ab} \p \bar u^{ab}- \half g_{ab}\p f^{ab}
-g\p f -\half(\bar\eta\p\bar\xi +\bar\xi\p\bar\eta) 
-\half \p\bar\phi\p\bar\phi, $$
$$S=  g,\quad S^{ab} = \bar u^{ac}\bar u^{bd} g_{cd}
-\bar u^{ab} g,\quad
S_a^b = \bar u^{bc} g_{ac} - \half \d_a^b g,
\quad S_{ab} = g_{ab},$$
where $\bar\g=\bar\eta e^{\bar\phi}$ and $\bar\b=\p\bar\xi e^{-\bar\phi}$. 
It is straightforward to
use the free field OPE's
\eqn\freeonebar{\bar v_{ab}(y) \bar u^{cd}(z) \to \d^c_{[a}\d^d_{b]} 
(y-z)^{-1},\quad
\bar\eta(y)\bar\xi(z)\to (y-z)^{-1},\quad \bar\phi(y)\bar\phi(z) 
\to -\log(y-z),}
$$g(y) f(z)\to (y-z)^{-1},\quad g_{ab}(y) f^{cd}(z) \to 
\d^c_{[a}\d^d_{b]} 
(y-z)^{-1},$$
to show that these currents satisfy the OPE's of \opebar.

\vskip 15pt
{\bf Acknowledgements:} I would like to thank Nikita Nekrasov and Warren
Siegel for extremely useful discussions
concerning the non-minimal variables and the regularization
factor, CNPq grant 300256/94-9, Pronex
grant 66.2002/1998-9, and FAPESP grant 04/11426-0 for partial financial
support, and the workshops Simons Workshop in Mathematics and Physics
at the YITP at SUNY at Stony Brook, and Mathematical Structures in
String Theory Workshop at the KITP at U.C. Santa Barbara for their
hospitality and partial financial support.

\listrefs
\end